\documentclass[twoside,12pt]{report}
\usepackage{graphicx}
\usepackage{amssymb, amsmath, amsxtra}
\usepackage{epsfig,subfigure}

\usepackage{latexsym}
\usepackage{bm}

\begin{document}
\begin{center}
{\huge \textbf{Nonlinear self-adjointness and conservation
laws}}\\[1ex]
 Nail H. Ibragimov\\
 Department of Mathematics and Science, Blekinge Institute
 of Technology,\\ 371 79 Karlskrona, Sweden
 \end{center}

 \noindent
\textbf{ Abstract.} The general concept of nonlinear
 self-adjointness of differential equations is introduced.
  It includes the linear self-adjointness as a particular case. Moreover, it
 embraces the strict self-adjointness and quasi self-adjointness introduced earlier by the author.
 It is shown that the equations
 possessing the nonlinear self-adjointness can be
 written equivalently in a strictly self-adjoint form by using appropriate
 multipliers. All linear equations possess the property of nonlinear
 self-adjointness, and hence can be rewritten in a nonlinear strictly self-adjoint.
 For example, the heat equation $u_t - \Delta u = 0$  becomes strictly self-adjoint after
 multiplying by $u^{-1}.$ Conservation laws associated with symmetries can be constructed
 for all differential equations and systems having the property of
 nonlinear self-adjointness.\\



  \section{Preliminaries}
 \label{nsa.pre}

  The concept of self-adjointness of nonlinear equations was introduced \cite{ibr06a}
 for constructing conservation laws associated with symmetries of differential
 equations. To extend the possibilities of the new method for constructing
 conservation laws \cite{ibr07a} the notion of quasi
 self-adjointness was used in \cite{ibr07d}.
 I suggest here the general concept of \textit{nonlinear self-adjointness}.
It embraces the strict self-adjointness and quasi self-adjointness
introduced earlier as well as the usual linear self-adjointness.
 Moreover, it will be shown that \textit{all linear equations}
 possess the property of nonlinear
 self-adjointness.

 It is demonstrated  that the equations
 possessing the nonlinear self-adjointness can be
 written equivalently in a strictly self-adjoint form by using appropriate
 multipliers. Consequently, any linear equation can be rewritten in an equivalent nonlinear form
which is strictly self-adjoint.
 For example, the heat equation $u_t - \Delta u = 0$  becomes strictly self-adjoint
 if we rewrite it in the form  $u^{-1}(u_t - \Delta u) = 0.$

 The construction of conservation laws demonstrates a practical significance
 of the nonlinear
 self-adjointness. Namely,
 \textit{conservation laws can be associated with symmetries for all linear
 and nonlinear self-adjoint differential equations}.

 \subsection{Notation}

Let $x = (x^1, \ldots, x^n)$  be independent variables. We
consider two sets of dependent variables, $u = (u^1, \ldots, u^m)$
and $v = (v^1, \ldots, v^m)$ with respective partial derivatives
 $$
 u_{(1)} = \{u^\alpha_i\}, \quad u_{(2)} = \{u^\alpha_{ij}\}, \
 \ldots, \   u_{(s)} = \{u^\alpha_{i_1\cdots i_s}\}
 $$
 and
 $$
 v_{(1)} = \{v^\alpha_i\}, \quad v_{(2)} = \{v^\alpha_{ij}\}, \
 \ldots, \   v_{(s)} = \{v^\alpha_{i_1\cdots i_s}\},
 $$
 where
 $$
 u^\alpha_i = D_i (u^\alpha)\,,
 \quad u^\alpha_{ij} = D_i D_j
  (u^\alpha)\,,  \quad u^\alpha_{i_1\cdots i_s}
  = D_{i_1} \cdots D_{i_s}(u^\alpha),
 $$
  $$
 v^\alpha_i = D_i (v^\alpha)\,,
 \quad v^\alpha_{ij} = D_i D_j
  (v^\alpha)\,,  \quad v^\alpha_{i_1\cdots i_s}
  = D_{i_1} \cdots D_{i_s}(v^\alpha).
 $$
  Here and in what follows $D_i$ denotes the operator of total
  differentiation:
$$
 D_i= \frac{\partial}{\partial x^i}
 + u^\alpha_i\frac{\partial}{\partial u^\alpha}
 + v^\alpha_i\frac{\partial}{\partial v^\alpha}
 +u^\alpha_{ij}\frac{\partial}{\partial u^\alpha_j}
  +v^\alpha_{ij}\frac{\partial}{\partial v^\alpha_j}
 + u^\alpha_{ijk}\frac{\partial}{\partial u^\alpha_{jk}}
 + v^\alpha_{ijk}\frac{\partial}{\partial v^\alpha_{jk}}  + \cdots\,.
 $$

  \subsection{Adjoint equations}
 We will
consider systems of $m$ differential equations (linear or
non-linear)
 \begin{equation}
 \label{nsa.eq1}
 F_\alpha \big(x, u, u_{(1)}, \ldots, u_{(s)}\big) = 0, \quad
 \alpha = 1, \ldots, m,
 \end{equation}
 with $m$ dependent variables.
  The \textit{adjoint equations} to equations (\ref{nsa.eq1}) are written
 \cite{ibr06a}
 \begin{equation}
 \label{nsa.eq2}
 F^*_\alpha (x, u, v, u_{(1)}, v_{(1)}, \ldots, u_{(s)}, v_{(s)})  = 0, \quad
 \alpha = 1, \ldots, m,
 \end{equation}
 with the adjoint operator $F^*_\alpha$ defined by
 \begin{equation}
 \label{nsa.eq3}
 F^*_\alpha (x, u, v, u_{(1)}, v_{(1)}, \ldots, u_{(s)}, v_{(s)}) =
  \frac{\delta {\cal L}}{\delta u^\alpha}\,,
 \end{equation}
 where ${\cal L}$ the  \textit{formal Lagrangian} for equations
 (\ref{nsa.eq1}) given by
 \begin{equation}
 \label{nsa.eq4}
 {\cal L} = \sum_{\beta= 1}^m v^\beta F_\beta(x, u, u_{(1)}, \ldots, u_{(s)})
 \end{equation}
 and $\delta/\delta u^\alpha$ is the variational derivative
 $$
 \frac{\delta}{\delta u^\alpha} =
\frac{\partial}{\partial u^\alpha} + \sum_{s=1}^\infty (-1)^s
D_{i_1}\cdots D_{i_s}\,\frac{\partial}{\partial
u^\alpha_{i_1\cdots i_s}}\,\cdot
 $$
  For a linear equation $L[u] = 0$ the adjoint operator
 defined by (\ref{nsa.eq3}) is identical with the classical adjoint operator $L^*[v]$
 determined by the equation
 $
 v L[u]-uL^*[v]=D_i(p^i).
 $

 The adjointness of \textit{linear} operators $L$ is a \textit{symmetric
 relation}, namely $(L^*)^* = L.$ Nonlinear equations do not possess
 this property so that, in general, $(F^*)^* \not= F.$

  \subsection{Self-adjointness}

 A linear operator $L$ is said to be self-adjoint if $L^* = L.$ Then
 we also say that the equation $L[u] = 0$ is self-adjoint. Thus, the
 self-adjointness of a linear equation $L[u] = 0$ means that
 the adjoint equation $L^*[v] = 0$ coincides with $L[u] = 0$ upon
 the substitution
 \begin{equation}
 \label{nsa.eq5}
  v = u.
 \end{equation}
 This property has been extended to nonlinear equations  \cite{ibr06a} by the following
 definition.\\
 \textbf{Definition 1.} Equation (\ref{nsa.eq1})
 is \textit{self-adjoint} if the adjoint equation
 (\ref{nsa.eq2}) becomes equivalent to the original equation (\ref{nsa.eq1})
 upon the substitution (\ref{nsa.eq5}).

 For example, the Korteweg-de Vries (KdV) equation
 $
 u_t = u_{xxx} + u u_x
 $
 is self-adjoint. Indeed, its adjoint equation (\ref{nsa.eq2}) has the form
 $
 v_t = v_{xxx} + u v_x
 $
 and coincides with the KdV equation  upon setting $v = u.$

 The concept of quasi self-adjointness introduced in \cite{ibr07d}
  generalizes Definition 1 by replacing
  (\ref{nsa.eq5}) with the substitution of the form
 \begin{equation}
 \label{nsa.eq6}
  v = \varphi(u), \quad \varphi'(u)  \not= 0.
 \end{equation}
 Thus, equation (\ref{nsa.eq1})
 is \textit{quasi self-adjoint} if the adjoint equation
 (\ref{nsa.eq2}) becomes equivalent to equation (\ref{nsa.eq1})
 upon the substitution (\ref{nsa.eq6}).  Let us consider as an example the equation
 \begin{equation}
 \label{nsa.eq7}
 u_t - u^2 u_{xx} = 0
 \end{equation}
 describing the nonlinear heat conduction in solid hydrogen \footnote{It is recalled that
 (\ref{nsa.eq7})  is related to the classical 1+1-dimensional heat equation by a
 differential substitution \cite{ibr85} or a reciprocal transformation \cite{rog85}.
 This connection, together with its extensions, allows the analytic solution of
certain moving boundary problems in nonlinear heat conduction
\cite{rog86}.}. Its
 adjoint equation (\ref{nsa.eq2}) is
 $$
 v_t + 4 u v u_{xx} + u^2 v_{xx} + 4u u_x v_x + 2 v u_x^2 = 0.
 $$
 It becomes equivalent to equation (\ref{nsa.eq7}) not after the
  substitution (\ref{nsa.eq5}) but after the following substitution
 of the form (\ref{nsa.eq6}):
  \begin{equation}
 \label{nsa.eq8}
   v  = u^{-2}\,.
 \end{equation}

 \subsection{Theorem on conservation laws}
 \label{tcl}

 We will use the following statement proved in \cite{ibr07a}.\\
 \textbf{Theorem 1.} Any infinitesimal symmetry (Lie point, Lie-B\"{a}cklund, nonlocal)
 $$
 X = \xi^i(x, u, u_{(1)}, \ldots) \frac{\partial}{\partial x^i} +
 \eta^\alpha (x,u, u_{(1)}, \ldots) \frac{\partial}{\partial u^\alpha}
 $$
 of equations (\ref{nsa.eq1}) leads to a conservation law $D_i (C^i) = 0$
 constructed by the formula
  \begin{equation}
 \label{nsa.eq9}
  C^i  =\xi^i {\cal L}+W^\alpha\,
 \Big[\frac{\partial {\cal L}}{\partial u_i^\alpha} -
 D_j \Big(\frac{\partial {\cal L}}{\partial u_{ij}^\alpha}\Big)
 + D_j D_k\Big(\frac{\partial {\cal L}}{\partial u_{ijk}^\alpha}\Big) -
 \ldots\Big]
  \end{equation}
 $$+D_j\big(W^\alpha\big)\,
 \Big[\frac{\partial {\cal L}}{\partial u_{ij}^\alpha} -
 D_k \Big(\frac{\partial {\cal L}}{\partial
 u_{ijk}^\alpha}\Big) + \ldots\Big]
 + D_j D_k\big(W^\alpha\big)\Big[\frac{\partial {\cal L}}{\partial u_{ijk}^\alpha} -
 \ldots\Big],
 $$
 where $ W^\alpha = \eta^\alpha - \xi^j u_j^\alpha$ and ${\cal L}$
 is the formal Lagrangian (\ref{nsa.eq4}). In applying the formula
 (\ref{nsa.eq9}) the formal Lagrangian ${\cal L}$ should be written
 in the symmetric form with respect to all mixed derivatives $u^\alpha_{ij}, \ u^\alpha_{ijk}, \ldots\,.$

 \section{Definition and main properties of nonlinear self-adjointness}
 \label{nsa.main}

  \subsection{Heuristic discussion}
   \label{nsa.main:1}

 Definition (\ref{nsa.eq4}) of the formal Lagrangian ${\cal L}$
 shows that the vector (\ref{nsa.eq9}) involves the `non-physical'
 variable $v.$ Therefore the validity of the conservation equation
 $D_i (C^i) = 0$ requires that we should take into account not only
 equations (\ref{nsa.eq1}) but also the adjoint equations (\ref{nsa.eq2}).
 But if the system (\ref{nsa.eq1}) is quasi self-adjoint (in
 particular self-adjoint), one can eliminate $v$ via the
 substitution (\ref{nsa.eq6}) and obtain a conservation law for equations
 (\ref{nsa.eq1}).

 However, the quasi self-adjointness is not the only case when the
 variables $v$ can eliminated from the conserved vector
 (\ref{nsa.eq9}). Let us note first of all that we can relax the condition
 $\varphi'(u)  \not= 0$ in (\ref{nsa.eq6}) since it is used only to
  guarantee the \textit{equivalence} of  equation
 (\ref{nsa.eq2}) to equation (\ref{nsa.eq1}) after eliminating $v$ by setting
  $v = \varphi(u).$ In constructing conservation laws, it is important only that
  $v$ does not vanish identically, because otherwise ${\cal L} = 0$ and
   (\ref{nsa.eq9}) gives the trivial vector $C^i = 0.$  Therefore
 we can replace condition $\varphi'(u)  \not= 0$ (\ref{nsa.eq6}) with the
 weaker condition $\varphi(u)  \not= 0.$ Secondly, the substitution (\ref{nsa.eq6})
 can be replaced with a more general substitution
 where $\varphi$
 involves not only the variable $u$ but also its derivatives as well
 as the independent variables $x.$ This will be a differential
 substitution
 \begin{equation}
 \label{nsa.eq10}
  v^\alpha = \varphi^\alpha(x, u, u_{(1)}, \ldots), \quad \alpha = 1, \ldots, m.
 \end{equation}
 The only requirement is that not all $\varphi^\alpha$ vanish.
 Moreover, $\varphi$ may involve nonlocal variables, e.g. $D_i^{-1}
 (u^\alpha).$ Then it is more convenient to determine $v$ implicitly
 by
 \begin{equation}
 \label{nsa.eq11}
 \Phi_\alpha(x, u, u_{(1)}, \ldots; v, v_{(1)}, \ldots) = 0, \quad \alpha = 1, \ldots, m.
 \end{equation}

  \subsection{Definition}
  \label{nsa.main:2}

  In this paper we will consider the
  substitutions (\ref{nsa.eq10}) that do not involve the derivatives
  and use the following definition of \textit{nonlinear
  self-adjointness}.\\
  \textbf{Definition 2.}  The system (\ref{nsa.eq1})
 is said to be \textit{self-adjoint} if the adjoint
 system (\ref{nsa.eq2}) is satisfied for all solutions $u$
 of equations (\ref{nsa.eq1}) upon a substitution
  \begin{equation}
  \label{nsa.eq12}
 v^\alpha = \varphi^\alpha(x, u), \quad \alpha = 1, \ldots, m,
 \end{equation}
such that
  \begin{equation}
  \label{nsa.eq13}
 \varphi(x, u) \not= 0.
 \end{equation}
 In other words,  the following equations hold:
  \begin{equation}
  \label{nsa.eq14}
 F^*_{\alpha} \big(x, u, \varphi(x, u), \ldots, u_{(s)}, \varphi_{(s)}\big) =
 \lambda^\beta_\alpha \, F_{\beta}\big(x, u,
 \ldots, u_{(s)}\big), \quad \alpha = 1, \ldots, m,
  \end{equation}
  where $\lambda^\beta_\alpha$ are undetermined
 coefficients. Here $\varphi$ is the $m$-dimensional vector $\varphi = (\varphi^1, \ldots, \varphi^m)$
  and $ \varphi_{(\sigma)}$ are its derivatives,
 $$
  \varphi_{(\sigma)}
  = \{D_{i_1} \cdots D_{i_\sigma}\big(\varphi^\alpha(x, u)\big)\}, \quad \sigma = 1, \ldots, s.
  $$
  Eq. (\ref{nsa.eq13}) means that not all components $\varphi^\alpha(x, u)$ of the vector  $\varphi$ vanish
 simultaneously.

  \subsection{Properties}
  \label{nsa.main:3}

 \textbf{Proposition 1.}  The system (\ref{nsa.eq1}) is self-adjoint in the sense of Definition 2 if
 and only of there exist functions $v^\alpha$ given by
 (\ref{nsa.eq12}) and satisfying the condition (\ref{nsa.eq13})
 that solve the adjoint system (\ref{nsa.eq2})
 \textit{for all solutions} $u(x)$ of equations (\ref{nsa.eq1}).\\
 \textbf{Proposition 2.}  Any linear equation  is self-adjoint.\\
 \textbf{Proof.} This is a consequence of the fact that the adjoint equation $L^*[v] = 0$ to the
 linear equation $L[u] = 0$ does not involve the variable $u.$
 Therefore any non-vanishing solution $v = \varphi (x)$ of the adjoint equation gives a vector function (\ref{nsa.eq12})
 which is independent of $u$ and hence satisfies the requirement of
 Definition 2.

 In the case of one dependent variable, i.e. $m = 1,$ we can easily prove the following.\\
 \textbf{Proposition 3.}  Equation (\ref{nsa.eq1}) is self-adjoint in the sense of Definition 2 if
 and only if it  becomes self-adjoint in the
 sense of Definition 1 upon rewriting in the
 equivalent form
 \begin{equation}
 \label{nsa.eq15}
 \mu (x, u) F\big(x, u, u_{(1)}, \ldots, u_{(s)}\big) = 0,
\quad  \mu (x, u) \not=  0,
 \end{equation}
 with an appropriate multiplier $\mu (x, u).$ In particular, any linear equation can be made
 self-adjoint in the sense of Definition 1.\\
 \textbf{Proof.} The computation reveals the following relation between the multiplier (\ref{nsa.eq15})
 and the substitution (\ref{nsa.eq12}):
  \begin{equation}
  \label{nsa.eq16}
 \varphi(x, u) = u \mu (x, u).
 \end{equation}
 Namely, if equation (\ref{nsa.eq1}) is self-adjoint in the sense of Definition
 2 with the substitution (\ref{nsa.eq12}), then equation
 (\ref{nsa.eq15}) whose multiplier $\mu$ is determined by
 (\ref{nsa.eq16}) is self-adjoint in the restricted sense of
 Definition 1, and visa versa.

  \subsection{Example: the Kompaneets equation}
  \label{nsa.main:4}

 Let us write the Kompaneets equation \cite{kom56}
  in the form
 \begin{equation}
 \label{nsa.eq17}
 u_t = \frac{1}{x^2}\, D_x\left[x^4 (u_x + u + u^2)\right].
 \end{equation}
 The reckoning shows that the adjoint equation to equation
 (\ref{nsa.eq17}),
 $$
 v_t + x^2 v_{xx} -x^2(1 + 2 u)v_x + 2 (x + 2 x u - 1)v = 0,
 $$
 does not have a solution of the form (\ref{nsa.eq6}) but it has the
 solution of the form (\ref{nsa.eq12}), namely $$v = x^2.$$ Hence,
 equation (\ref{nsa.eq17}) is not quasi self-adjoint, but it is
 self-adjoint in the sense of Definition 2. Equation
 (\ref{nsa.eq16}) provides the multiplier $\mu = x^2/ u.$ Hence the
 Kompaneets equation becomes self-adjoint in the sense of
 Definition 1 if we write it in the form
 \begin{equation}
 \label{nsa.eq18}
 \frac{x^2}{u}\,u_t = \frac{1}{u}\,D_x\left[x^4 (u_x + u +
 u^2)\right].
 \end{equation}

 \section{Time-dependent conservation laws for the KP equation}
  \label{nsa.KP}


 We will use  the KP equation (Kadomtsev-Petviashvili \cite{KP70})
 \begin{equation}
 \label{nsa.eq19}
 u_{tx} - u u_{xx} - u^2_x - u_{xxxx} = u_{yy}
 \end{equation}
 written as the system (see, e.g. \cite{nov80}, p. 241, and the
 references therein)
 \begin{equation}
 \label{nsa.eq20}
 u_t - u u_x - u_{xxx} - \omega_y = 0, \quad \omega_x - u_y = 0.
 \end{equation}

 \subsection{Self-adjointness}

 The formal Lagrangian (\ref{nsa.eq4}) for equations (\ref{nsa.eq20}) is written
 \begin{equation}
 \label{nsa.eq21}
 {\cal L} = v (u_t - u u_x - u_{xxx} - \omega_y) + z(\omega_x -
 u_y)
 \end{equation}
  and equation (\ref{nsa.eq3}) yields the following adjoint system to the system (\ref{nsa.eq20}):
 \begin{equation}
 \label{nsa.eq22}
 v_t - u v_x - v_{xxx} - z_y = 0, \quad z_x - v_y = 0.
 \end{equation}
 Equations (\ref{nsa.eq22}) become identical with the KP
 equations (\ref{nsa.eq20}) upon the substitution
 \begin{equation}
 \label{nsa.eq23}
 v = u, \quad z = \omega.
 \end{equation}
 It means that the system (\ref{nsa.eq20}) is self-adjoint.

 \subsection{Symmetries}

  The system (\ref{nsa.eq20}) admits the infinite-dimensional Lie algebra spanned by the operators
 \begin{equation}
 \label{nsa.eq24}
 \begin{split}
 & X_f =3 f\frac{\partial}{\partial t}
  + (f'x+\frac{1}{2}\,f''y^2)\frac{\partial}{\partial x}
  + 2f'y \frac{\partial}{\partial y}\\[1 ex]
   & -\big[2 f'u + f''x + \frac{1}{2}\,f'''y^2\big]\frac{\partial}{\partial u}
   - \big[3 f' \omega + f'' y u + f''' xy
  +\frac{1}{6} f^{(4)}y^3\big]\frac{\partial}{\partial \omega}\,,
  \end{split}
   \end{equation}
 \begin{equation}
 \label{nsa.eq25}
  X_g = 2 g\frac{\partial}{\partial y} + g'y \frac{\partial}{\partial x}
   - g''y \frac{\partial}{\partial u} -
   \big[g' u + g''x+\frac{1}{2}\, g'''y^2\big]\frac{\partial}{\partial \omega}\,,
 \end{equation}
 \begin{equation}
 \label{nsa.eq26}
 X_h = h \frac{\partial}{\partial x} - h' \frac{\partial}{\partial u}
 - h'' y \frac{\partial}{\partial \omega}\,,
 \end{equation}
 where $f, g, h$ are three arbitrary functions of $t.$
 We will ignore the obvious symmetry
 $$
 X_\alpha = \alpha(t) \frac{\partial}{\partial  \omega}
 $$
 describing the addition to $\omega$ an
 arbitrary function of $t.$

 Note, that the operators (\ref{nsa.eq24})-(\ref{nsa.eq26}) considered without the
 term $\frac{\partial}{\partial \omega}$ span the infinite-dimensional Lie algebra of
 symmetries of the KP equation (\ref{nsa.eq19}). They coincide (up to normalizing coefficients)
 with the symmetries of the KP equation that were first
 obtained by F. Schwarz in 1982 (see also \cite{dklw85},
 \cite{ame-rog89}  and the references therein).

 \subsection{Conservation laws}

 Noether's theorem is not applicable to the system
 (\ref{nsa.eq20}). But Theorem 1 from Section \ref{tcl} is applicable.
 Applying formula (\ref{nsa.eq9}) to the formal Lagrangian
 (\ref{nsa.eq21}) and to the symmetry (\ref{nsa.eq24}), then eliminating  $v,
 z$ by the substitution (\ref{nsa.eq23}) we obtain the
 conservation law
 \begin{equation}
 \label{nsa.eq27}
 \left[D_t (C^1) + D_x (C^2) + D_y
 (C^3)\right]_{(\ref{nsa.eq20})} = 0
 \end{equation}
 with the following components of the conserved vector $C = (C^1,
 C^2, C^3):$
 \begin{equation}
 \label{nsa.eq28}
 \begin{split}
 C^1 & =  - \frac{1}{2}\, f' u^2 -
  \left(x f'' + \frac{1}{2}\,y^2 f'''\right) u,\\[1ex]
 C^2 & = \left(u u_{xx} + \frac{1}{3}\,u^3 - \frac{1}{2}\,u_x^2 - \frac{1}{2}\, \omega^2\right) f'
 + \left(x u_{xx} + \frac{1}{2}\, x u^2 - u_x\right) f''\\[1ex]
 & + \frac{1}{4}\, \left(y^2 u^2 + 2 y^2 u_{xx} - 4 x y \omega\right) f'''
 - \frac{1}{6}\, y^3 \omega f^{(4)},\\[1ex]
 C^3 & = u \omega f' + x \omega f''
 +  \left(x y u + \frac{1}{2}\, y^2\omega \right) f'''
  + \frac{1}{6}\, y^3 u f^{(4)}.
  \end{split}
\end{equation}
 The  conservation equation (\ref{nsa.eq27}) for the vector
 (\ref{nsa.eq28}) has the form
 \begin{equation}
 \label{nsa.eq29}
 \begin{split}
 & D_t(C^1) + D_x(C^2) + D_y(C^3)\\[1ex]
  & = \left(u f' + x f'' + \frac{1}{2}\, y^2f'''\right)
 (u_{xxx} + u u_x + \omega_y - u_t)\\[1ex]
 & + \left(\omega f' + xy f''' + \frac{1}{6}\, y^3 f^{(4)}\right)
 (u_y - \omega_x).
 \end{split}
 \end{equation}
 Since $f = f(t)$ is an arbitrary function, (\ref{nsa.eq28}) provides an infinite set
 of conserved vectors. Note that the subscript
 (\ref{nsa.eq20}) in Eq. (\ref{nsa.eq27}) refers to restriction on the solution manifold of Eqs.
 (\ref{nsa.eq20}).
  I did not find the conserved vector (\ref{nsa.eq28}) with arbitrary $f(t)$ in previous
 publications, e.g. in \cite{oev-fuc81}. The symmetries (\ref{nsa.eq25}) and (\ref{nsa.eq26}) lead
  to the conserved vectors
   \begin{equation}
 \label{nsa.eq29A}
 \begin{split}
 C^1 & =  y u g'',\\[1ex]
 C^2 & = \left(x \omega - y u_{xx} - \frac{1}{2}\,y u^2\right) g''
 + \frac{1}{2}\, y^2 \omega g''',\\[1ex]
 C^3 & = - (x u + y \omega) g'' - \frac{1}{2}\, y^2 u g'''
  \end{split}
 \end{equation}
 and
  \begin{equation}
 \label{nsa.eq29B}
 \begin{split}
 C^1 & =  u h',\\[1ex]
 C^2 & = y \omega h'' - \left(u_{xx} + \frac{1}{2}\, u^2\right) h',\\[1ex]
 C^3 & = - \omega h' - y u h'',
  \end{split}
 \end{equation}
 respectively.

 \section{Conservation laws for linear equations}
  \label{nsa.con:lin}

 One can obtain conserved vector by formula (\ref{nsa.eq9}) for any linear equation
 because linear equations are self adjoint according to Proposition 2. Consider, e.g. the heat equation
 \begin{equation}
 \label{nsa.eq30}
 u_t - \Delta u = 0
 \end{equation}
with any number of spatial variables $x = (x^1, \ldots, x^n).$
 Applying formula (\ref{nsa.eq9}) to $ X = u \frac{\partial}{\partial u}$
 we obtain the conservation law
 $\left[D_t(\tau) + \nabla \cdot \chi\right]_{(\ref{nsa.eq30})} = 0$
 with
 \begin{equation}
 \label{nsa.eq31}
 \tau = \varphi (t, x) u, \quad \chi = u \nabla \varphi (t, x) - \varphi (t, x) \nabla u,
 \end{equation}
 where $v = \varphi (t, x)$ is an arbitrary solution of the adjoint equation $v_t + \Delta v = 0$ to
 equation (\ref{nsa.eq30}). The conserved vector (\ref{nsa.eq31})
 embraces the conserved vectors associated with all other symmetries
 of equation (\ref{nsa.eq30}). In particular, the projective
 symmetry
 $$
 X = t^2 \frac{\partial}{\partial t} + t x^i \frac{\partial}{\partial x^i}
 - \frac{|x|^2 + 2 n t}{4}\, u \frac{\partial}{\partial u}
 $$
 of equation (\ref{nsa.eq30}) gives the conserved vector
 $$
 \tau = \frac{|x|^2 - 2 n t}{4}\, u, \quad \chi^i = \frac{x^i}{2}\,u  - \frac{|x|^2 - 2 n t}{4}\,
 u_i\,.
 $$
 It corresponds to (\ref{nsa.eq31}) with the particular solution
 $v = (|x|^2 - 2 n t)/4$ of the adjoint equation.  In one-dimensional
 case ($n = 1$), it is shown in \cite{dor-svir83} by direct calculation that all conserved vectors for the
 heat equation $u_t = u_{xx}$ have the form (\ref{nsa.eq31}). See
 also \cite{ste-wol81}.



\begin{thebibliography}{10}

 \bibitem{ibr06a}
 Ibragimov N H  2006 Integrating factors, adjoint equations and
 Lagrangians  \textit{J. Math. Anal. Appl.} \textbf{318} 742--57

 \bibitem{ibr07a}
 Ibragimov N H  2007 A new conservation theorem
 \textit{J. Math. Anal. Appl.} \textbf{333} 311--28

 \bibitem{ibr07d}
  Ibragimov N H  2007 Quasi self-adjoint differential equations
 \textit{Archives of ALGA} \textbf{4} 55--60

 \bibitem{ibr85}
 Ibragimov N H  1983 \textit{Transformation groups applied to mathematical physics}
  (Moscow: Nauka). English transl. Dordrecht: Reidel 1985

 \bibitem{rog85}
  Rogers C 1985 Application of a reciprocal transformation to a two phase Stefan problem \textit{J.
 Phys. A: Math. Gen.} \textbf{18} L105-L109

 \bibitem{rog86}
 Rogers C 1986 On a class of moving boundary value problems in nonlinear heat conduction:
 application of a {B}\"{a}cklund transformation \textit{Int. J.
 Nonlinear Mechanics} \textbf{21} 249-256

 \bibitem{kom56}
 Kompaneets A S 1956 The establishment of thermal equilibrium
 between quanta and electrons \textit{Zh. Eksp. Teor. Fiz} \textbf{31}
 876--85\\
  Kompaneets A S 1957 \textit{Sov. Phys. - JETP} \textbf{4}  730 (Engl. Transl.)

 \bibitem{KP70}
 Kadomtsev B B and  Petviashvili V I 1970
 On stability of solitary waves in weakly dispersive media
 \textit{Dokl. Akad. Nauk SSSR} \textbf{192} 753--56\\
  Engl. Transl.: Kadomtsev B B and  Petviashvili V I 1970  \textit{Sov. Phys. Dokl.}
   \textbf{15} 539-42

 \bibitem{nov80}
 Novikov S, Manakov S V, Pitaevskii L P and Zakharov V E 1984
 \textit{Theory of solitons: The inverse scattering method}
 (New York: Consultants Bureau)

 \bibitem{dklw85}
 David D,  Kamran N, Levi D and Winternitz P 1985 Subalgebras
 of loop algebras and symmetries of the
 Kadomtsev-Petviashvili equation
 \textit{ Phys. Rev. Letters} \textbf{55} 2111 -2113

\bibitem{ame-rog89}
 Rogers C and Ames W F 1989 \textit{Nonlinear boundary value problems
 in science and engineering}  (Boston: Academic Press)

 \bibitem{oev-fuc81}
 Oevel W and Fuchsssteiner B 1982
 Explicit formulas for symmetries and conservation laws of the
 Kadomtsev-Petviashvili equation \textit{Phys. Letters}
 \textbf{88A} 323--27

 \bibitem{dor-svir83}
Dorodnitsyn V A  and Svirshchevskii S R 1983 On Lie-B\"acklund
groups admitted by the heat equation with a source
\textit{Preprint 101} (Moscow: Inst. Appl. Math. USSR Acad. Sci.)

 \bibitem{ste-wol81}
Steinberg S  and Wolf K B 1981 Symmetry, conserved quantities and
moments in diffusive equations \textit{J. Math. Anal. Appl.}
\textbf{80} 36--45

\end{thebibliography}
 \end{document}